\documentclass[pra,aps,amsmath,twocolumn,showpacs,floatfix]{revtex4-1}
\usepackage{graphicx}
\usepackage{bm}
\usepackage{pdfpages}
\input{epsf}
\begin{document}
\epsfverbosetrue
\def\la{{\langle}}
\def\ra{{\rangle}}
\def\vep{{\varepsilon}}
\newcommand{\beq}{\begin{equation}}
\newcommand{\eeq}{\end{equation}}
\newcommand{\beqa}{\begin{eqnarray}}
\newcommand{\eeqa}{\end{eqnarray}}
\newcommand{\q}{\quad}
\newcommand{\h}{\hat{H}}
\newcommand{\ha}{\hat{h}}
\newcommand{\p}{\partial}
\newcommand{\A}{|\Omega'|}
\newcommand{\AC}{{\it AC}}
\newcommand{\n}{\\ \nonumber}
\newcommand{\om}{\omega}
\newcommand{\Om}{\Omega}
\newcommand{\os}[1]{#1_{\hbox{\scriptsize {osc}}}}
\newcommand{\cn}[1]{#1_{\hbox{\scriptsize{con}}}}
\newcommand{\sy}[1]{#1_{\hbox{\scriptsize{sys}}}}
\title{A comment on the paper "How can a Result of a Single Coin Toss Turn Out to be 100 Heads" by C. Ferrie and J. Combes}
\affiliation{$^b$ IIKERBASQUE, Basque Foundation for Science, Maria Diaz de Haro 3, 48013, Bilbao, Spain}
\begin{abstract}
\end{abstract}
\author {D. Sokolovski$^{a,b}$}
\affiliation{$^a$ Departamento de Qu\'imica-F\'isica, Universidad del Pa\' is Vasco, UPV/EHU, Leioa, Spain}
\affiliation{$^b$ IIKERBASQUE, Basque Foundation for Science, Maria Diaz de Haro 3, 48013, Bilbao, Spain}

\begin{abstract}
The authors of  a recent paper [Phys. Rev. Lett. \textbf{113}, 120404 (2014)] suggest that "weak values are not inherently quantum but rather a purely statistical feature of 
pre- and postselection with disturbance". We argue that this claim is erroneous,
since such values require averaging with distributions which change sign. This type of averaging  arises naturally in quantum mechanics, 
but may not occur in classical statistics.
\end{abstract}
\maketitle
\vskip0.5cm

%
%
%
\section {Introduction}
One needn't read much beyond the title of \cite{PRL} to see that the answer to its question must be 'It cannot'. No matter how elaborate the protocol, some Alice at the end of the line wold receive the coin and write down the value of $1$ or $-1$, depending on whether the coin shows up heads or tails . Adding these numbers, and dividing them by the number of trials $N$, would always yield a value between $-1$ and $1$. So what is wrong with the argument of \cite{PRL}?

\section{'normal' and 'anomalous' averages }
Consider an average of the form 
\begin{eqnarray}\label{1}
\bar{s}=\sum_{n=1}^N s_n P_n,\q \sum_n P_n=1
\end{eqnarray}
where $s_1>s_2...>s_N$. We will call $\bar{s}$ {\it normal} if it lies between $s_1$ and $s_N$, 
$s_1\ge \bar{s} \ge s_N$, and {\it anomalous} otherwise. It is readily seen that 
$\bar{s}$ is always normal if $P_n\ge 0$, $n=1,2,..N$, and to be anomalous it requires that 
at least one of the $P$'s is negative. To see how an anomalous average may be produced, consider
$N=2$, $s_{1,2}=\pm1$, $P_1=1001$, $P_2=-1000$, $P_1+P_2=1$, and find that
$\bar{s}=2001$. Thus, a large anomalous value would occur where the moduli of $P_n$ are large, 
but the sum of all $P_n$ is unity due to a very precise cancellation.
Multiplying the $P_n$'s by $s_n$ destroys the cancellation, so that the resulting $\bar{s}$  is unduly large.
A more detailed example is given in \cite{ANN}, where a similar effect is found responsible for what appears
as super-luminal transmission of a wave packet. 
So where else is one likely to encounter anomalous mean values?
\newline
\section{Quantum weak values}
The place to look is in quantum mechanics.
Consider a two level quantum system (a spin $1/2$)  with a hamiltonian $H$, pre- and post-selected (observed) in some states 
$\psi$ and $\phi$ at $t=0$ and $t=T$, respectively. Choose an operator $S$ with eigenstates $|s_{1,2}\ra$ and eigenvalues
$\pm1$. Inserting, at some $0<t'<T$ the unity $I=|s_{1}\ra\la s_1|+|s_{2}\ra\la s_2|$ into the transition amplitude
$\la \phi|\exp(-iHT)|\psi\ra$,  shows that the spin can reach the final state via two virtual 
routes, $\psi\to s_1 \to \phi$  and $\psi \to s_2 \to \phi$. Putting for simplicity $H=0$, for the two corresponding amplitudes we 
have
\begin{eqnarray}\label{1a}
A_{1,2}=\la \phi |s_{1,2}\ra\la s_{1,2}|\psi\ra.
\end{eqnarray}
To see what actually happens at $t'$
we may employ a von Neumann pointer with position $f$, 
initially 
 decoupled from the spin.
 We set the pointer at some $f'$ by preparing in in a  state $|M_{f'}\ra=\int df G(f-f') |f\ra$, where $G(f)$ is a function 
 peaked around $0$ with a width $\Delta f$, such that $\int|G(f)|^2=1$.
  For  $t' <t<t+\tau$ the pointer briefly interacts with the spin via
$H_{int}=-i\tau^{-1}\partial_f A$, and then its final position is measured exactly. 
 Now $\Delta f $ determines what we know about the initial position of the pointer and, therefore, the accuracy of the measurement.
For $\Delta f<<1$, the accuracy is good, and the interference between two routes is destroyed. With $f'=0$ he pointer's readings are narrowly grouped around the values $f=\pm 1$
which occur with the probabilities $P_{1,2}=|A_{1,2}|^2/(|A_{1}|^2+|A_{2}|^2)$, respectively.
Thus, the average of $S$ at $t'$ is given by
\begin{eqnarray}\label{2}
\bar{s}_{acc}(\phi|\psi)=(|A_{1}|^2-|A_{2}|^2)/(|A_{1}|^2+|A_{2}|^2),
\end{eqnarray}
With $P_{1,2}\ge 0$ it is always normal, and lies between $-1$ and $1$.
\newline
Our measurement can be made less accurate it two different ways. A {\it classical} uncertainty
is added when we still have an accurate meter which destroys interference between the two 
routes, but cannot, for some reason, set it exactly to zero. With its initial
position being $f'$ with a probability $W(f')=W(-f')$, peaked  around $0$ with a width $\delta f'$, the initial meter state is a mixture, 
\begin{eqnarray}\label{3}
\rho_M=\int|M_{f'}\ra W(f')\la|M_{f'}|df',   \q \int W(f')df'=1.
\end{eqnarray}
If $\delta f'$ is large, the final pointer's readings are widely spread, and we must associate a reading of $100$ not with 
the actual value of the spin's component, but rather with the shortcomings of our measurement. 
Note that after averaging over many trials we still find that the mean pointer reading is given by Eq.(\ref{2}).
In other words, as far as the averages are concerned, we still have the same information, 
it is just harder to obtain.
\newline
The other possibility is more interesting.
 A {\it quantum} uncertainty can be introduced by preparing the meter in a pure state (\ref{3}) and selecting $\Delta f\to \infty$.
 We still do not know the pointer's initial position, but in a quite different sense: there is no probability for it to be $f'$. 
Now the pointer's readings are spread over a range $\sim \Delta f$, and it still seems unwise to associate the 
value of $100$ with an intrinsic property of the spin. 
The central result of \cite{AHAR} (see also \cite{ME}) is that
 the mean pointer reading is given not by (\ref{2}),  but by the real part of an expression similar to  (\ref{2}), with  $|A_n|^2$  
replaced by the amplitudes themselves
\begin{eqnarray}\label{4}
\bar{s}_{weak}(\phi|\psi)=Re\{(A_{1}-A_{2})/(A_{1}+A_{2})\}=Re \frac{\la \phi|A|\psi\ra}{\la \phi|\psi\ra}.\q
\end{eqnarray}
Note that since the final meter's readings $f$ are distributed between $-\infty$ and $\infty$ with non-negative 
probability density $p(f)=\ge 0$,  there is nothing strange about $\bar{s}_{weak}$ taking any real value, however large.
It can, alternatively, be seen as an average of the variable $s$ numbering our two routes, and taking the values $+1$ and 
$-1$, $\bar{s}_{weak}(\phi|\psi)=\sum_{n=1,2}s_n P_n$, with $P_{1,2}=Re \{A_{1,2}/(A_1+A_2)\}$.
Since $A_1$ and $A_2$ are arbitrary complex quantities, $P_{1,2}$ may or not be non-negative, and 
the average $\bar{s}_{weak}$ may be both normal or anomalous. For example, it is always normal 
for $\phi=\psi$ and $A_{1,2}=|\la s_{1,2}|\psi\ra|^2\ge 0$, or anomalous, e.g., for  for $\phi=\psi$, or $100$ for $ReA_{1,2}=0$, $A_2/A_1=-99/101$.
Thus, initial meter's position is uncertain in the quantum sense, $\bar{s}_{weak}(\phi|\psi)$ no longer gives one an indication that we started with 
exactly two routes labelled $1$ and $-1$. The question is now of of interpreting this result, and here we disagree with the authors of \cite{AHAR}.
\section{Anomalous values and the uncertainty principle}
The answer goes back to one of basic postulates of quantum mechanics. A very inaccurate 'weak' measurement 
does not destroy interference between the two routes  \cite{ME}. Accordingly, there are only the amplitudes, and not the probabilities, assigned to the two 
virtual routes, and used in Eq.(\ref{4}).
Further, the Uncertainty Principle states that two interfering routes cannot be told apart and should 
be considered a single pathway \cite{Feyn}. If so, the question {\it "what was, on average, the value of $S$ if we hadn't destroyed coherence between the two routes?"} should not have meaningful answer \cite{ME}. Although a mean value of $100$ for a spin 1/2 looks like a wrong result produced by a malfunctioning meter, yet it cannot simply be 'corrected', since the 'correct' answer just doesn't exist. This is a purely quantum dilemma: do we ascribe the same degree of importance and 'reality'  to the results of the accurate and the inaccurate 'weak' measurements, just because weak measurements can  and have been made.
It is worth arguing in favor of accurate 'strong' measurements.
For any choice of $\psi$ and $\phi$, an accurate pointer would only yield the values $1$ and $-1$. This is how one knows that spin of $1/2$ is an intrinsic property of the electron, 
and is later able to  write its wave function as a Pauli spinor in situations which has nothing to do with the original Stern-Gerlach experiment.
One who only has access to the averages (\ref{2}), but not the distributions, may at least note that for any $\psi$ and $\phi$ $\bar{s}_{strong}(\phi|\psi)$ lies between $-1$ and $1$.
This is still a property of the electron, and not just of the chosen transition. But from $\bar{s}_{weak}(\phi|\psi)$ one can only deduce that, depending on $\psi$ and $\phi$, 
the expectation value of $S$ may take an arbitrary value. Such value is a property of the particular transition, and does not have a more general meaning.
The answer to a question which should not have an answer is, in this case,  {\it "anything at all"}.
\section{Where did Ferrie and Combes go wrong?}
A classical theory operates with strictly non-negative probabilities which are ascribed to all 
observable scenarios from the very beginning.
 Consequently, all 
observable averages must be of the normal type, and lie between the smallest and the largest values the variable 
of interest may take.
The authors of \cite{PRL} claim to have demonstrated the contrary using a simple classical model, but their analysis contains an error. 
The classical model in \cite{PRL} has two initial states $\psi$, two final states $\phi$, and two intermediate states $s$, all labelled by $\pm1$.
With the choice $\psi=1$ and $\phi=-1$ there are two pathways $1\to \pm 1 \to -1$ travelled with the probabilities [cf. Eq.(27) of Ref.\cite{PRL}]
\begin{eqnarray}\label{5}
P_{1,2}=(1\pm\lambda-\delta)/2(1-\delta).
\end{eqnarray}
Here $0<\lambda<1$ is a small parameter, and $\delta$ is chosen so that $P_{1,2}$ are strictly positive, 
$0<\delta<1-\lambda$. With this choice the authors find that
\begin{eqnarray}\label{6}
\overline{s/\lambda}=\sum_{n=1,2}(-1)^nP_n/\lambda=1/1-\delta
\end{eqnarray}
can be made arbitrarily small, provided $\lambda$ is small, and $\delta$ is close to unity.
This, claim the authors of Ref.\cite{PRL}, is an anomalous weak value obtained in a purely 
classical context, and this is where they are wrong. The mean in (\ref{6}) is the mean not of the $s$ taking the values of $\pm1$, but of the
variable $s/\lambda=\pm1/\lambda$, and $1 <\overline{s/\lambda} < 1/\lambda$ is a perfectly normal average, 
well within the allowed interval  $[-1/\lambda,1/\lambda]$. The same criticism applies to the 'classical protocol' illustrated in Fig.1, and described below the Eq.(34),
since it relies on Eqs.(\ref{5}) and (\ref{6})
\newline
To put it plain, with $\lambda =1/200$ the authors of \cite{PRL} tell Alice-at-the-end-of-the-line to write down 
$200$ if he/she receives the coin showing up heads, or $-200$ otherwise. 
Adding the numbers and dividing by the number of trials may now yield a value of $100$, 
which is then shown to the reader as the proof that there is 'a simple classical model which exhibits anomalous weak values' \cite{PRL}. What the authors of \cite{PRL} should have been saying, in answering their own question, is:
'It can, if you call 'heads' - '200 heads', and 'tails' - '200 tails'. But this observation doesn't add much to our understanding of weak measurements or weak values .

\section{Conclusions}
In summary, appearance of anomalous weak values is a strictly quantum phenomenon.
Such values arise in a situation where the Uncertainty Principle forbids gaining required 
information about a quantum system under observation. 
Since for an average to be anomalous the 'probabilities' in Eq.(\ref{1}) must change sign,
anomalous weak values cannot arise for any observable quantity in 
classical statistics.

\end{document}